\documentclass[journal]{IEEEtran}
\usepackage[T1]{fontenc}
\usepackage{graphicx}
\usepackage{lettrine}
\usepackage{cite}
\usepackage{booktabs}
\usepackage{amsmath}
\usepackage{amsthm}
\usepackage{array}
\usepackage{bm}
\usepackage{algorithm}
\usepackage{amssymb}
\usepackage{algpseudocode}
\usepackage{flushend}
\usepackage{times}
\usepackage{hhline}
\usepackage{siunitx}
\usepackage{xcolor}
\usepackage[dvipsnames]{xcolor}
\usepackage{subcaption}

\newtheorem{theorem}{Theorem}

\let\origtau\tau
\renewcommand{\tau}{\scalebox{1.4}{$\origtau$}}

\usepackage[switch,columnwise]{lineno}

\begin{document}

\title{Net-Zero Power UAVs: Facilitating Quantum HRLLC in Multi-Hop UAV Repeaters}

\author{Ali~Ranjha,
        Georges Kaddoum,~\IEEEmembership{Senior Member, IEEE},
         and Halim Yanikomeroglu~\IEEEmembership{Fellow, IEEE}
\thanks{
A. Ranjha and H. Yanikomeroglu are with the Department of Systems and Computer Engineering, Carleton University, Ottawa, Canada, e-mails: AliRanjha@cunet.carleton.ca, ali-nawaz.ranjha.1@ens.etsmtl.ca and halim@sce.carleton.ca, respectively.\\
G. Kaddoum is with the Department of Electrical Engineering, \'Ecole de Technologie Sup\'erieure, Montr\'eal, Canada, e-mail: Georges.Kaddoum@etsmtl.ca
}}  
       
\maketitle

\begin{abstract}

In this paper, we propose a novel joint optimization framework for autonomous, quantum-enabled UAV repeater networks designed to support hyper-reliable low-latency communications (HRLLC) in challenging environments such as post-explosion chemical plant operations. The proposed system uses a symmetric dual-UAV architecture with three sequential quantum channels operating under strict energy harvesting (EH) constraints. We formulate a highly challenging non-convex mixed-integer non-linear programming (MINLP) problem. This model jointly determines qubit allocation, 2-D uncrewed aerial vehicle (UAV) spatial positioning (distance and altitude), and transmit powers to minimize the total decoding error.
Crucially, our framework explicitly models real-world quantum physics, including decoherence, gate imperfections, and entanglement degradation. Since the objective function is demonstrably non-convex, to efficiently solve the MINLP, we propose a quantum-resource-optimized GA (QRO-GA). The results of our numerical simulations show that the QRO-GA yields better results than benchmark methods and it is highly efficient, with the convergence speed significantly increasing with the total qubit allocation. Convergence is achieved in as few as 5 iterations for higher qubit allocation. Moreover, the results reveal a critical U-shaped tradeoff, proving that an optimal latency distribution within a strict $80\ \mu$s transmission window is strictly required to maintain HRLLC reliability. In addition, we find that the QRO-GA successfully optimizes the 2-D UAV positioning to minimize total decoding error. Finally, it balances consumed and harvested power to achieve net-zero power UAVs system.

\end{abstract}

\vspace{-5mm}
\section{Introduction}
\IEEEPARstart{Q}{uantum-enabled} uncrewed aerial vehicle (UAV) communication systems represent a paradigm shift in aerial vehicular communications for mission-critical applications in challenging environments that lack conventional infrastructure. 
Integrating quantum communication principles with UAV-based repeater networks offers fundamental advantages in information-theoretic security and data fidelity for vehicular connectivity \cite{gisin2002quantum}. This is particularly important for hyper-reliable low-latency communications (HRLLC) that require decoding error rates below $10^{-9}$. However, free-space quantum communication is inherently sensitive to atmospheric turbulence, noise, and decoherence \cite{liu2020drone}. In complex terrains with obstructed or degraded direct line-of-sight (LoS), standard single-hop quantum links can easily fail. An unprecedented solution to
overcome this physical fragility is quantum-enabled UAVs operating as coordinated multi-hop aerial repeaters provide. By dynamically optimizing their symmetric positions and qubit allocations, these aerial repeaters reduce effective channel lengths and actively mitigate atmospheric degradation, thus maintaining reliable connectivity during critical operations.

Previous literature on UAV-assisted vehicular communications has primarily addressed classical systems. On the performance analysis front, rigorous statistical frameworks were established for classical UAV-assisted optical relays. These frameworks included comprehensive closed-form evaluations of bit error rates and outage probabilities for dual-hop free-space optical (FSO) systems under atmospheric turbulence \cite{xu2023outage}, as well as asymmetric RF/FSO space-air-ground integrated networks \cite{qu2022uav}. Meanwhile, on the resource allocation front, previous approaches focused on trajectory optimization \cite{dong2025attention,im2025trajectory}, power allocation \cite{qiu2025joint}, and anti-jamming techniques \cite{wen2025joint} using successive convex approximations (SCA). However, while these classical frameworks extensively modelled environmental fading and standard routing, their limitation is that they fundamentally neglected quantum-specific phenomena such as decoherence effects, entanglement degradation, quantum noise, and gate imperfections that dominate real-world quantum hardware performance in aerial platforms.
On the theoretical quantum communication front, while channel capacity bounds have been established \cite{kondamuri2024quantum}, yet these frameworks remain disconnected from practical UAV deployment under operational constraints, including propulsion power, mobility limitations, and energy harvesting inherent to aerial vehicles.

Yet another critical research gap persists in vehicular quantum communications: the joint optimization of quantum resource (qubit) allocation across multiple aerial repeater channels while simultaneously determining optimal UAV positioning and transmit powers under power harvesting constraints remains unexplored. Moreover, computational complexity characterization of such joint optimization for aerial vehicular systems has not been established. This gap is particularly pronounced for symmetric UAV configurations where geometric coupling between aerial vehicle position and channel quality creates non-trivial optimization landscapes.

To address these gaps, in this study, we propose a quantum-enabled dual-UAV repeater framework for HRLLC. First, quantum physical layer dynamics offer fundamental security advantages, thus ensuring the ultra-low decoding error rates ($<10^{-9}$) that are vital for post-disaster environments while eliminating classical RF jamming vulnerabilities. However, deploying quantum links in aerial networks introduces several severe physical challenges: qubits are highly fragile to atmospheric decoherence over distance, and UAVs are strictly bound by net-zero energy harvesting requirements, which creates a complex mixed-integer optimization landscape. To overcome these concerns, the proposed symmetric dual-UAV architecture physically shortens transmission hops to mitigate decoherence. Finally, we propose a Quantum-Resource-Optimized Genetic Algorithm (QRO-GA) to solve the resulting non-convex MINLP, jointly optimizing qubit allocation, 3D positioning, and transmit powers to achieve strict net-zero operation and bounded latency while minimizing total decoding error.
\vspace{-5mm}
\section{System Model and Problem Formulation}
\begin{figure}[!t]
\centering
\includegraphics[height=4.0cm,width=8.2cm]{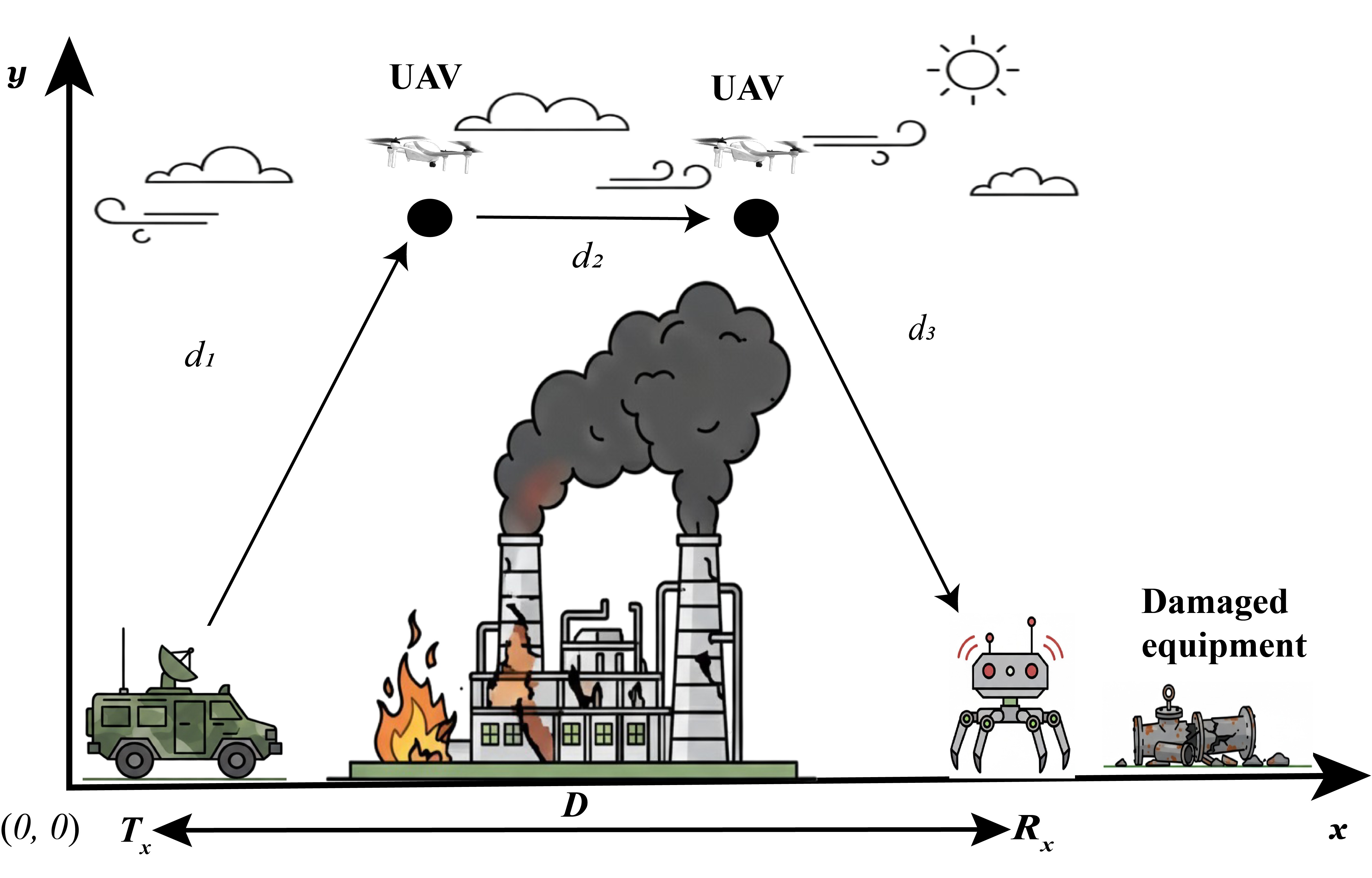}
\caption{Illustration of the dual-UAV repeater system with energy harvesting for HRLLC.}
\label{fig:system_model}
\end{figure}

We consider a quantum communication system deployed for post-explosion chemical plant remediation, where HRLLC is critical for remote operation. The system comprises a command station ($T_{x}$), two quantum-enabled UAV repeater nodes, and a remediation robot ($R_{x}$), (see Figure~\ref{fig:system_model}) to fix damaged equipment. Damaged plant structures, debris, and dense smoke plumes preclude direct communication between the command station and the robot, necessitating aerial quantum repeaters. While the operational objective is the delivery of classical command and control data for the remediation robot, the physical layer utilizes quantum carriers to ensure security and reliability in the degraded environment. Specifically, classical HRLLC bits are encoded into the quantum states of photonic qubits, which are then transmitted over optical channels.\ The system uses symmetric UAV positioning where both repeater nodes operate at a common altitude $z \in [z_{\min}, Z_{\max}]$. The first UAV is located at a horizontal distance $x \in [0, X_{\max}]$ from $T_{x}$, while the second UAV is symmetrically positioned at distance $x$ from $R_{x}$ or equivalently, distance $D-x$ from the $T_{x}$. This creates three sequential quantum channels: $T_{x}$ to the first UAV, the first UAV to the second UAV, and, finally, the second UAV to the $R_{x}$ robot. The symmetric geometry ensures balanced channel conditions for the first and third hops, while the inter-UAV link distance varies with $x$, thus allowing for dynamic optimization around local obstacles. The communication distances for the three channels are as follows:
\begin{align}
d_1^2(x, z) &= x^2 + z^2, \label{eq:d1}\\
d_2^2(x) &= (D - 2x)^2, \label{eq:d2}\\
d_3^2(x, z) &= x^2 + z^2, \label{eq:d3}
\end{align}
where the symmetric design ensures $d_1(x,z) = d_3(x,z)$. Here, free-space channel gains follow the path loss shown below:
\begin{equation}
h_i(x, z) = \frac{\beta_0}{d_i^2(x, z)}, \quad i \in \{1, 2, 3\},
\end{equation}
where $\beta_0$ is the channel gain constant at unit distance. While dense smoke introduces particulate scattering, our architecture assumes that UAVs operate at altitudes ($z \ge z_{\min}$) that are sufficient to clear ground-level macro-structures, thus maintaining a dominant LoS. Considering the severe non-convexity introduced by integrating quantum decoherence with net-zero energy constraints, we adopt a free-space path-loss model to establish a tractable theoretical baseline, thereby deferring complex stochastic fading models to future extensions.

\vspace{-5mm}
\subsection{Quantum Error Model}
We assume $M$ denotes total available qubits distributed as $\bm{m} = [m_1, m_2, m_3]$ satisfying $\sum_{i=1}^{3} m_i = M$ with $m_i \geq 1$. The quantum noise parameter from Heisenberg uncertainty is $\delta_q = \sigma_q \hbar f_q$, where $\sigma_q$ is noise level, $\hbar = h/(2\pi)$ is reduced Planck constant, and $f_q$ is qubit frequency. The effective SNR for channel $i$ with transmit power $P_i$ is as follows:
\begin{equation}
\gamma_i(x, z) = \frac{P_i h_i(x, z)}{1 + \delta_q}.
\end{equation}

The quantum measurement variance is derived by scaling the classical Gaussian channel dispersion \cite{polyanskiy2010channel} with quantum hardware constraints \cite{nielsen2010quantum}, yielding the following:
\begin{equation}
V_i(x, z) = \left[1 - (1 + \gamma_i(x, z))^{-2}\right] \eta_q \xi,
\end{equation}
Physically, the bracketed term represents the fundamental information variance bounded by the effective SNR, identical to classical AWGN dispersion. However, in a quantum framework, this ideal dispersion is scaled by quantum gate efficiency denoted by $\eta_q$ and entanglement preservation factor given by $\xi$. These parameters account for the physical reality where imperfect quantum measurements and decoherence further degrade the observable information density.
 Now, channel capacity under decoherence \cite{cai2013algebraic} can be computed as follows:
\begin{equation}
C_i(m_i, x, z) = \log_2(1 + \gamma_i(x, z)) \exp\left(-\rho \frac{m_i}{f_q}\right),\label{eq:CapacityDecoherence}
\end{equation}
where $\rho$ is decoherence rate. Importantly, (\ref{eq:CapacityDecoherence}) models the classical data rate over quantum channels, explicitly accounting for throughput reduction due to qubit decoherence. Thereafter, the error exponent for payload length $L$  \cite{polyanskiy2010channel} is by the following:
\begin{equation}
f_i(m_i, x, z) = \ln(2) \sqrt{\frac{m_i}{V_i(x, z)}} \left[C_i(m_i, x, z) - \frac{L}{m_i}\right].
\end{equation}
Here, the per-channel error probability is as follows:
\begin{equation}\small
\varepsilon_i(m_i, x, z) = \begin{cases}
Q(f_i) + \sigma_q \sqrt{m_i} \exp(-m_i \eta_q), & f_i > 0, \\
0.5 + |f_i|/2, & f_i \leq 0,
\end{cases}
\end{equation}\normalsize
where $Q(\cdot)$ is the Gaussian Q-function. Physical realizability requires
\begin{equation}\small
\max\left\{\frac{\hbar\rho}{2 m_i h f_q}, \varepsilon_i(m_i, x, z)\right\} \leq 0.5,
\end{equation}\normalsize
where the first term is the Heisenberg limit \cite{zhou2018achieving}. Now, the total decoding error is given by the following:
\begin{equation}\small
\varepsilon_{\text{tot}}(\bm{m}, x, z, \bm{P}) = \sum_{i=1}^3 \varepsilon_i(m_i, x, z).
\end{equation}\normalsize
\vspace{-10mm}
\subsection{Energy Harvesting Model}
Each UAV harvests solar and wind power at altitude $z$ \cite{lee2017optimal}. The solar power is modelled as:
\begin{equation}
S(z) = \max\left\{A_{\text{solar}} \phi(z) C(z), 0\right\},
\end{equation}
where $\phi(z) = \alpha - \beta_s e^{-z/\delta}$ models atmospheric attenuation, while
\begin{equation}\small
C(z) = \begin{cases} 
e^{-\beta_c (L_{\text{up}} - L_{\text{low}})}, & z \leq L_{\text{low}}, \\
e^{-\beta_c (L_{\text{up}} - z)}, & L_{\text{low}} < z \leq L_{\text{up}}, \\
1, & z > L_{\text{up}}
\end{cases}\normalsize
\end{equation}
represents cloud layer effects.
Moreover, the wind power \cite{sekander2020statistical} is computed as follows:
\begin{equation}\small
W(z) = \begin{cases} 
0, & v(z) \leq V_{ci}, \\
\frac{1}{2} \rho_w A C_p v(z)^3, & V_{ci} < v(z) \leq V_r, \\
P_r, & V_r < v(z) \leq V_{co}, \\
0, & v(z) > V_{co},
\end{cases}\normalsize
\end{equation}
where $v(z) = v_0 (z/z_0)^{1/7}$ is the wind speed profile. Combining solar and wind powers we obtain a total harvested power: $P_{\text{harvest}}(z) = S(z) + W(z)$. Importantly, both solar and wind power models are non-linear in nature.

\subsection{Latency Model}
To meet HRLLC requirements \cite{chen2021urllc}, the end-to-end (E2E) delay for each multi-hop repeater link $i \in \{1, 2, 3\}$ should satisfy the following:
\begin{equation}
    D_i = \tau_i + T_{\text{HARQ}} + T_{\text{proc}} \leq T_{\text{th}},
\end{equation}
where $\tau_i = m_i / B_{\text{sym}}$ is the transmission delay for $m_i$ qubits, $T_{\text{HARQ}}$ is the round-trip propagation delay, and $T_{\text{proc}}$ is the processing delay. By isolating $\tau_i$, the maximum allowable transmission duration becomes $\tau_{\max} = T_{\text{th}} - (T_{\text{HARQ}} + T_{\text{proc}})$. Consequently, the latency budget strictly bounds the per-hop blocklength, as shown in Eq. \ref{eq:blocklength_bound}:
\begin{equation}
    m_i \leq \lfloor \tau_{\max} \cdot B_{\text{sym}} \rfloor \triangleq m_{\text{max\_lat}}.
    \label{eq:blocklength_bound}
\end{equation}
\subsection{Optimization Problem}
We aim to minimize the total decoding error by jointly optimizing the UAV coordinates $(x, z)$, qubit distribution $\bm{m}$, and transmit power $\bm{P}$:
\small
\begin{align}
\textbf{P1:} \quad \min_{\bm{m}, x, z, \bm{P}} \quad & \varepsilon_{\text{tot}}(\bm{m}, x, z, \bm{P}) \label{eq:obj}\\
\text{s.t.} \quad 
& \sum_{i=1}^{3} m_i = M, \tag{C1}\\
& m_i \in \mathbb{Z}^+, \quad 1 \leq m_i \leq m_{\text{max\_lat}}, \quad \forall i \in \{1, 2, 3\}, \tag{C2}\\
& 0 \leq x \leq X_{\max}, \tag{C3}\\
& z_{\min} \leq z \leq Z_{\max}, \tag{C4}\\
& z \geq x \tan(\theta_{\min}), \tag{C5}\\
& \sum_{i=1}^{3} P_i + 2P_{\text{UAV}} \leq P_{\text{harvest}}(z), \tag{C6}\\
& P_{\min} \leq P_i \leq P_{\max}, \quad \forall i \in \{1, 2, 3\}. \tag{C7}
\end{align}
\normalsize

Constraints C1-C2 define the latency-bounded qubit allocation, C3-C5 establish the geometric and stability boundaries, C6 enforces the system-wide net-zero energy operation \footnote{We use inequality ($\leq$) rather than strict equality to model an onboard battery buffer. Strict equality unrealistically dictates the instantaneous consumption of all harvested energy, thus rendering transient energy surpluses mathematically infeasible. By contrast, the inequality correctly permits excess energy to be stored, ensuring sustained net-zero operation.}, and, finally, C7 bounds the individual transmission powers.

\section{Quantum-Resource Optimized Genetic Algorithm (QRO-GA)}


\textbf{P1} is a non-convex Mixed-Integer Non-Linear Programming (MINLP) problem (see Appendix) containing non-differentiable terms and integer variables, rendering classical gradient solvers ineffective. To overcome this, we propose a Qubit Resource Allocation Genetic Algorithm (QRO-GA) that encodes our seven decision variables into a chromosome $\bm{c}$ to minimize the decoding error fitness function $\mathcal{F}(\bm{c})$, natively bypassing local optima without gradient reliance.

\begin{algorithm}[!t]
\caption{Quantum-Resource-Optimized GA (QRO-GA)}
\begin{algorithmic}[1]
\State \textbf{Input:} Pop. size $N_{\text{pop}}$, Max gens $G_{\text{max}}$, parameters $p_c, p_m$, $N_{\text{elite}}$
\State \textbf{Output:} Optimal decision vector $\bm{c}^*$
\State Initialize population $\mathcal{P}^{(0)} = \{\bm{c}_1, \ldots, \bm{c}_{N_{\text{pop}}}\}$ uniformly within bounds
\State Evaluate fitness $\mathcal{F}(\bm{c}_i)$ for all $\bm{c}_i \in \mathcal{P}^{(0)}$ including constraint penalties
\For{$g = 1$ to $G_{\text{max}}$}
    \State $\mathcal{E}^{(g)} \gets$ Extract top $N_{\text{elite}}$ individuals from $\mathcal{P}^{(g-1)}$
    \State $\mathcal{O}^{(g)} \gets \emptyset$ \Comment{Initialize offspring pool}
    \While{$|\mathcal{O}^{(g)}| < N_{\text{pop}} - N_{\text{elite}}$}
        \State Select parents $\bm{c}_p, \bm{c}_q$ from $\mathcal{P}^{(g-1)}$ via tournament selection
        \State $\bm{c}_1, \bm{c}_2 \gets$ Apply SBX crossover to $\bm{c}_p, \bm{c}_q$ (with probability $p_c$)
        \State $\bm{c}_1, \bm{c}_2 \gets$ Apply polynomial/integer mutation (with probability $p_m$)
        \State Evaluate fitness $\mathcal{F}$ for $\bm{c}_1, \bm{c}_2$ and append to $\mathcal{O}^{(g)}$
    \EndWhile
    \State $\mathcal{P}^{(g)} \gets \mathcal{E}^{(g)} \cup$ Select best $(N_{\text{pop}} - N_{\text{elite}})$ from $\mathcal{O}^{(g)}$
    \If{$|\mathcal{F}^{(g)}_{\text{best}} - \mathcal{F}^{(g-1)}_{\text{best}}| < \tau$} 
        \State \textbf{break} \Comment{Early convergence met}
    \EndIf
\EndFor
\State \Return $\bm{c}^* = \arg\min_{\bm{c} \in \mathcal{P}^{(g)}} \mathcal{F}(\bm{c})$
\end{algorithmic}
\end{algorithm}

\begin{figure*}[!t]
    \centering
    
    \begin{minipage}{0.3\textwidth}
        \includegraphics[width=1.1\linewidth]{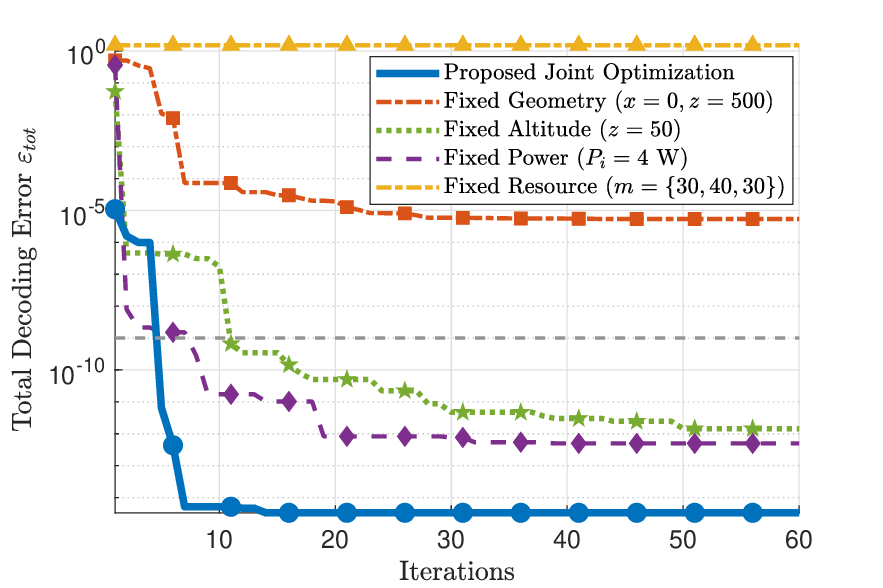}
        \caption{Convergence behavior of the proposed QRO-GA compared to benchmark methods.}
        \label{fig:benchmarks}
    \end{minipage}\hfill
    \begin{minipage}{0.3\textwidth}
        \includegraphics[width=1.1\linewidth]{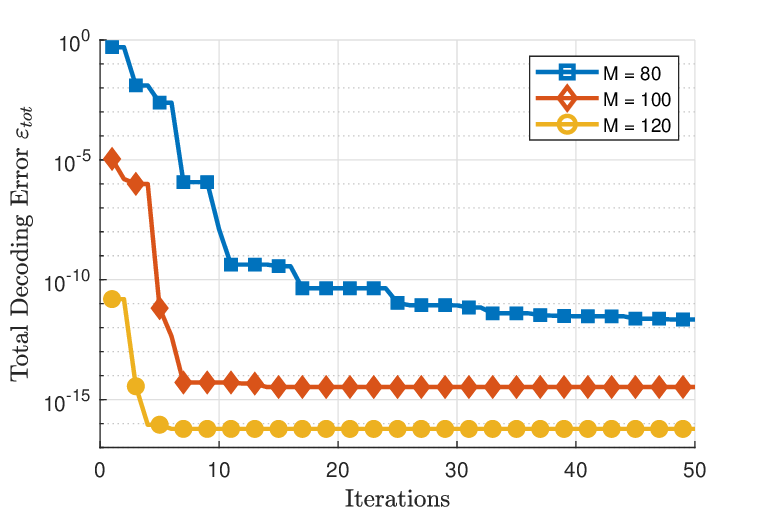}
        \caption{Convergence behavior of the proposed QRO-GA for different Qubit Resource Allocations.}
        \label{fig:convergence}
    \end{minipage}\hfill
    \begin{minipage}{0.3\textwidth}
        \includegraphics[width=1.1\linewidth]{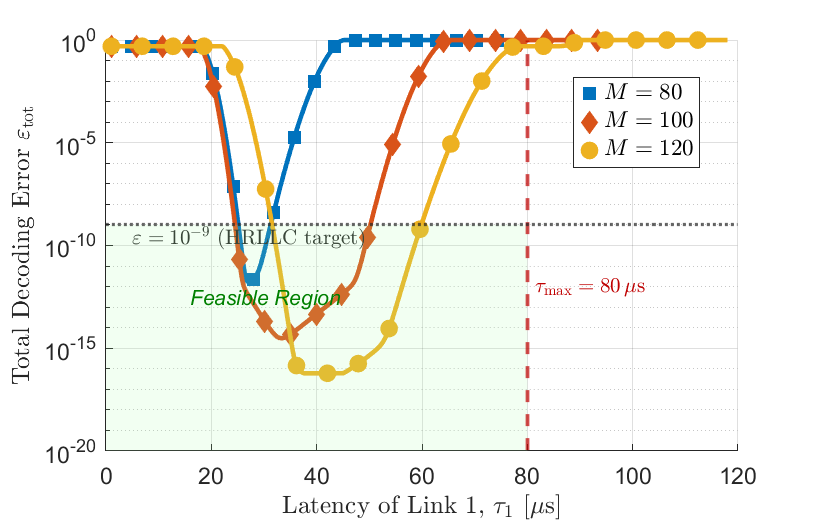}
        \caption{Total Decoding Error vs. Latency.}
        \label{fig:latency}
    \end{minipage}
    
    \vspace{0.6cm} 
    
    \begin{minipage}{0.3\textwidth}
        \centering
        \includegraphics[width=1.5\linewidth]{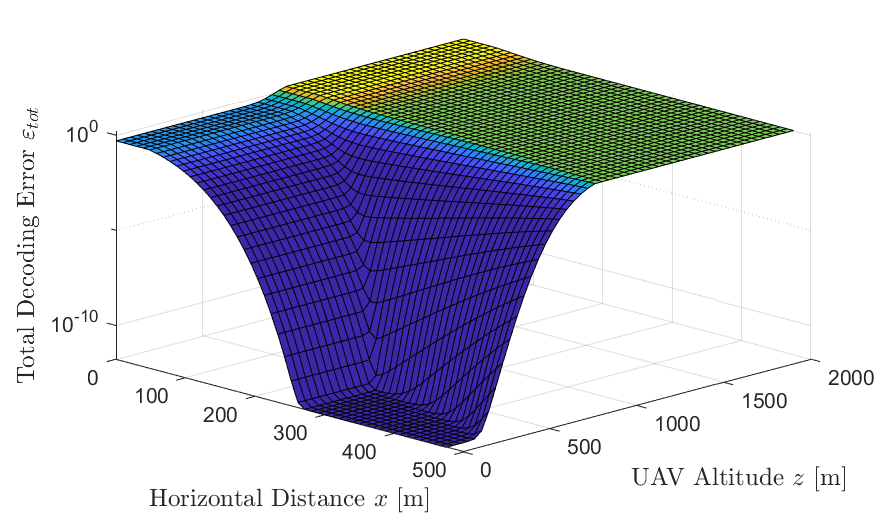}
        \caption{Total decoding error vs. 2-D positioning.}
        \label{fig:ErrorvsPositioning}
    \end{minipage}\hspace{0.05\textwidth} 
    \begin{minipage}{0.3\textwidth}
        \centering
        \includegraphics[width=1.5\linewidth]{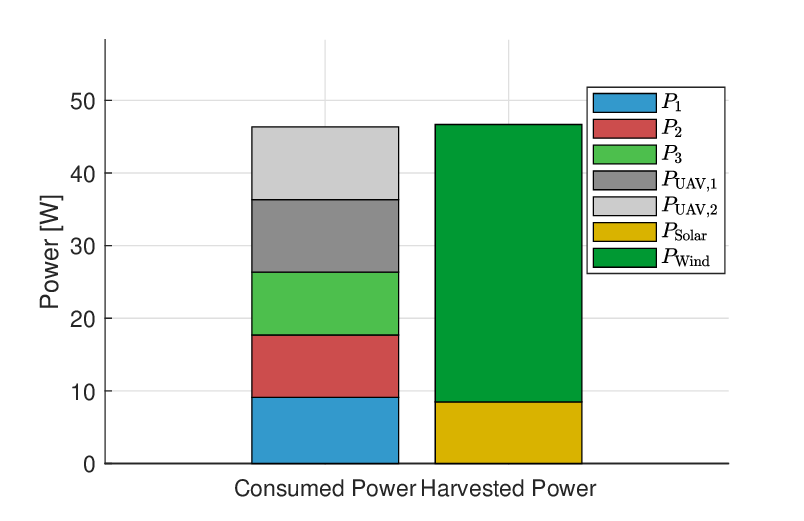}
        \caption{Consumed power vs. harvested power.}
        \label{fig:PowerProfile}
    \end{minipage}
    
\end{figure*}
\section{Simulation Results}

The proposed QRO-GA algorithm was configured with specific operational and environmental parameters to ensure reliable UAV-based quantum communication under geometric, energy, and atmospheric constraints. The UAVs were assumed to operate over a maximum horizontal distance $D = 1000$~m with a minimum safe separation of $D_{\mathrm{safe}} = 50$~m, resulting in a maximum horizontal displacement $X_{\max} = 475$~m. To satisfy line-of-sight requirements, the minimum elevation angle was set to $\theta_{\min} = 20^\circ$.  
The genetic algorithm utilized a population size $N_{\mathrm{pop}} = 200$, elite count $N_{\mathrm{elite}} = 15$, maximum generations $G_{\mathrm{max}} = 150$, and crossover fraction $p_c = 0.8$. Adaptive mutation was employed to maintain solution diversity and prevent premature convergence, with the remaining parameters set as shown in Table~I. Regarding computational cost, each generation entails fitness evaluation scaling as $O(N_{\text{pop}} \cdot T_{\text{eval}})$, where $T_{\text{eval}}$ is the cost of computing $\varepsilon_{\text{tot}}$, followed by selection at $O(N_{\text{pop}} \log N_{\text{pop}})$ and genetic operations at $O(N_{\text{pop}} \cdot n_{\text{vars}})$. Consequently, the total time complexity of the proposed scheme is dominated by the fitness evaluation, resulting in $O(G_{\text{max}} \cdot N_{\text{pop}} \cdot T_{\text{eval}})$.

\begin{table}[h!]\footnotesize
\centering
\caption{Key Operational and Environmental Parameters}
\setlength{\tabcolsep}{3pt} 
\begin{tabular}{lll | lll}
\hline
\textbf{Param} & \textbf{Value} & \textbf{Unit} & \textbf{Param} & \textbf{Value} & \textbf{Unit} \\
\hline
$D$ & 1000 & m & $L_{\mathrm{low}}$ & 500 & m \\
$D_{\mathrm{safe}}$ & 50 & m & $L_{\mathrm{up}}$ & 2000 & m \\
$X_{\max}$ & 475 & m & $v_0$ & 5 & m/s \\
$\theta_{\min}$ & 20 & deg & $z_0$ & 10 & m \\
$M$ & 100 & qubits & $V_{\mathrm{ci}}$ & 3 & m/s \\
$L$ & 100 & bits & $V_r$ & 12 & m/s \\
$f_q$ & $5 \times 10^9$ & Hz & $V_{\mathrm{co}}$ & 25 & m/s \\
$P_{\min}$ & 0.1 & W & $V_{\mathrm{stab}}$ & 15 & m/s \\
$P_{\max}$ & 10 & W & $\rho_w$ & 1.225 & kg/m$^3$ \\
$P_{\mathrm{UAV}}$ & 10 & W & $A$ & 0.5 & m$^2$ \\
$S_p$ & 1361 & W/m$^2$ & $P_r$ & 30 & W \\
\hline
\end{tabular}
\end{table}

First, Figure \ref{fig:benchmarks} evaluates the convergence behavior of the proposed QRO-GA against the following four optimization benchmarks: fixed geometry, fixed altitude, fixed power, and fixed resource allocation. We can observe that the fixed resource allocation instantly fails, while restricting physical variables (such as geometry or power) forces the system to stall at suboptimal solutions. By contrast, the proposed QRO-GA dynamically and simultaneously balances all interconnected variables, thus vastly outperforming the fixed baselines and achieving the best convergence well below the $\varepsilon = 10^{-9}$ HRLLC target. This unequivocally proves the need for the joint optimization framework.\
Thereafter, Figure \ref{fig:convergence} illustrates the convergence behavior of the proposed QRO-GA for various total qubit allocations ($M=80, 100, 120$). A clear trend is observed: with an increase of M, convergence speed significantly improves. Specifically, the lowest allocation ($M=80$) requires approximately 30 iterations to converge to its error floor. This convergence becomes substantially more rapid for higher allocations, achieved in only 10 iterations for $M=100$, and further accelerating to just 5 iterations for $M=120$. Furthermore, the final achievable Total Decoding Error ($\varepsilon_{\text{tot}}$) dramatically decreases, confirming that a larger quantum resource pool directly translates to a higher communication reliability.  
Now, Figure \ref{fig:latency} shows the total decoding error vs. the transmission latency of the first link ($\tau_1$). The U-shaped curves highlight the resource allocation tradeoff under strict total qubit budgets ($M$), demonstrating that an optimal distribution is strictly required to remain within the green HRLLC feasible region. This region is bounded by a strict transmission limit of $\tau_{\max} = 80\ \mu$s, which is derived by subtracting $20\ \mu$s of HARQ and hardware processing overhead from the $100\ \mu$s 3GPP factory automation target \cite{3gpp_ts_22_104}.\
Afterwards, Figure~\ref{fig:ErrorvsPositioning} shows the relationship between the Total Decoding Error ($\varepsilon_{\text{tot}}$) and the optimized 2-D UAV spatial positioning (horizontal distance $x$ and altitude $z$). This visualization clearly demonstrates the trade-offs involved in the joint optimization framework, highlighting the specific $[x, z]$ coordinates that are the most favorable for achieving the high reliability necessary to support HRLLC services. These results confirm that both the horizontal distance and altitude must be precisely optimized to minimize link degradation and ensure maximum communication performance.

Finally, Figure \ref{fig:PowerProfile} presents the system's energy balance by comparing the total consumed power against the total harvested power. The consumed power is the sum of all operational loads, including the following: the transmit powers for the three sequential quantum links ($P_1, P_2, P_3$) and the operational powers ($P_{\text{UAV,1}}$) and ($P_{\text{UAV,2}}$) required for hovering. The total harvested power ($P_{\text{harvest}}$) is the aggregate of both solar and wind energy captured at the optimized altitude. Crucially, Figure \ref{fig:PowerProfile} demonstrates that the total consumed power is equivalent to the total harvested power, which confirms that the energy constraint (C6) is strictly satisfied and establishing the operation of a net-zero power UAVs system.

\section{Conclusion}

In this paper, we investigated a joint optimization framework for quantum-enabled UAV repeater networks designed to support HRLLC under strict energy harvesting constraints. To this end, we formulated the complex interplay between quantum resource allocation (qubits), UAV 2-D spatial positioning, and transmit powers as a non-convex MINLP problem. To effectively solve this computationally challenging problem, we proposed a QRO-GA. Our numerical results confirmed the superior performance and efficiency of the QRO-GA, which significantly outperformed fixed benchmark methods. Specifically, the analysis revealed that , with an increase of the total qubit allocation ($M$),
the total decoding error ($\varepsilon_{\text{tot}}$) dramatically decreases achieving rapid convergence in as few as 5 iterations for higher allocations. Furthermore, the simulations exposed a critical U-shaped tradeoff, proving that an optimal latency distribution within a strict 80 $\mu$s transmission window is absolutely required to maintain HRLLC reliability. In addition, the algorithm successfully optimized the 2-D UAV spatial positioning to minimize the total decoding error. Finally, the QRO-GA determined optimal operating points that perfectly balanced consumed and harvested power, thus achieving a net-zero system and ensuring perpetual UAV flight while strictly satisfying the non-linear energy constraints.

\appendix 

\begin{theorem}[Non-Convex Objective]\label{thm:nonconvex_obj}
The objective function $\varepsilon_{\text{tot}}(\bm{m}, x, z, \bm{P})$ is non-convex over its domain.
\end{theorem}

\begin{proof}
We establish non-convexity by analyzing the Hessian of the channel gain $h_1(x,z) = \beta_0/(x^2 + z^2)$ as follows:
\begin{equation}
\bm{H}_{h_1} = \frac{\beta_0}{(x^2+z^2)^3} 
\begin{bmatrix}
6x^2 - 2z^2 & 8xz \\
8xz & 6z^2 - 2x^2
\end{bmatrix}.
\end{equation}
Computing $\text{tr}(\bm{H}_{h_1}) = 4\beta_0/(x^2+z^2)^2$ and $\det(\bm{H}_{h_1}) = -12\beta_0^2/(x^2+z^2)^4$, the eigenvalues are as follows:
\begin{equation}
\lambda_1 = \frac{6\beta_0}{(x^2+z^2)^2} > 0, \quad \lambda_2 = \frac{-2\beta_0}{(x^2+z^2)^2} < 0.
\end{equation}
Since the Hessian has both positive and negative eigenvalues, it is indefinite and $h_1(x,z)$ exhibits saddle-point geometry. Considering that $\varepsilon_{\text{tot}}$ is a composite function involving $h_1(x,z)$, the non-convexity propagates to the objective.
\end{proof}

\vspace{-5mm}
\bibliographystyle{IEEEtran}
\bibliography{Bibliography.bib}
\end{document}